\documentclass[showpacs,pra,superscriptaddress,10pt,twocolumn]{revtex4}%
\usepackage{amsfonts}
\usepackage{amsmath}
\usepackage{amssymb}
\usepackage{graphicx}%
\begin{document}
\title{Entanglement charge of thermal states}
\author{Ming-Yong Ye}
\affiliation{School of Physics and Optoelectronics Technology, Fujian Normal University,
Fuzhou 350007, China}
\author{Xiu-Min Lin}
\affiliation{School of Physics and Optoelectronics Technology, Fujian Normal University,
Fuzhou 350007, China}
\author{Yan-Kui Bai}
\affiliation{College of Physical Science and Information Engineering and Hebei Advance Thin
Films Laboratory, Hebei Normal University, Shijiazhuang, Hebei 050016, China}
\author{Z. D. Wang}
\affiliation{Department of Physics and Center of Theoretical and Computational Physics,
University of Hong Kong, Pokfulam Road, Hong Kong, China}

\begin{abstract}
Entanglement charge is an operational measure to quantify nonlocalities in
ensembles consisting of bipartite quantum states. Here we generalize this
nonlocality measure to single bipartite quantum states. As an example, we
analyze the entanglement charges of some thermal states of two-qubit systems
and show how they depend on the temperature and the system parameters in an
analytical way.

\end{abstract}

\maketitle

\textit{Introduction.}--It has been indicated that ensembles consisting of
bipartite or multipartite quantum states can exhibit a kind of nonlocality
that is different from quantum entanglement \cite{bennett}. So far, most
efforts have been devoted to address conditions that can be used to determine
whether a given ensemble can exhibit this nonlocality
\cite{ye,cc1,c1,c2,c3,c4,c5,c6,c7,4bell,2state,cohen}. Compared to the
numerous studies about quantum entanglement, profound understandings on this
intriguing ensemble nonlocality and its implications are still awaited. In a
recent paper, we have introduced "entanglement charge" as an operational
measure to quantify nonlocalities in ensembles consisting of bipartite quantum
states \cite{charge}. The quantification of nonlocalities in ensembles has
also been discussed in Ref. \cite{quanti ensem}.

In this paper we generalize the idea of entanglement charge to single
bipartite quantum states. The generalization that can be done is due to the
fact that quantum states have ensemble decompositions \cite{book}. Based on
the entanglement charge, two kinds of nonlocalities on bipartite states are
introduced, which are different from quantum entanglement. So the work can
enrich our knowledge about quantum states. The paper is organized as follows.
We first give a brief introduction to entanglement charge defined for
ensembles consisting of bipartite quantum states. Then we generalize the
concept of entanglement charge to single bipartite quantum states and apply it
to some thermal states. Finally, a summary is given.

\textit{Entanglement charge of ensembles.}--Suppose $\varepsilon=\left\{
p_{X},\rho_{X}^{AB}\right\}  $ is an ensemble consisting of bipartite states.
The entanglement charge $N\left(  \varepsilon\right)  $ of the ensemble
$\varepsilon$ may be positive, negative or zero \cite{charge}. The ensembles
with positive $N\left(  \varepsilon\right)  $ are defined to have information
nonlocality and those with negative $N\left(  \varepsilon\right)  $ are
defined to have entanglement nonlocality. In both cases the entanglement
charge $N\left(  \varepsilon\right)  $ or its absolute value $\left\vert
N\left(  \varepsilon\right)  \right\vert $ can be used as a measure to
quantify the corresponding nonlocality.

Usually it is hard to compute $N\left(  \varepsilon\right)  $. However, when
the states $\rho_{X}^{AB}$ in the ensemble $\varepsilon=\left\{  p_{X}%
,\rho_{X}^{AB}\right\}  $ are mutually orthogonal pure states, the
entanglement charge $N\left(  \varepsilon\right)  $ satisfies the following
bounds
\begin{align}
N\left(  \varepsilon\right)   &  \leq S\left(  A\left\vert B\right.  \right)
=S\left(  \rho^{AB}\right)  -S\left(  \rho^{B}\right)  ,\label{up}\\
N\left(  \varepsilon\right)   &  \leq S\left(  B\left\vert A\right.  \right)
=S\left(  \rho^{AB}\right)  -S\left(  \rho^{A}\right)  ,\label{up2}\\
N\left(  \varepsilon\right)   &  \geq\sum p_{X}S\left(  \rho_{X}^{A}\right)
-I_{\rho^{AB}}\left(  A;B\right)  , \label{low}%
\end{align}
where $\rho_{X}^{A}=Tr_{B}\rho_{X}^{AB}$, $\rho^{AB}=\sum_{X}p_{X}\rho
_{X}^{AB}$, $\rho^{B}=Tr_{A}\rho^{AB}$, $\rho^{A}=Tr_{B}\rho^{AB}$, $S\left(
\cdot\right)  $ is the quantum entropy and $I_{\rho^{AB}}\left(  A;B\right)
=S\left(  \rho^{A}\right)  +S\left(  \rho^{B}\right)  -S\left(  \rho
^{AB}\right)  $ is the quantum mutual information \cite{book}.

Especially, when $\rho_{X}^{AB}$ in $\varepsilon=\left\{  p_{X},\rho_{X}%
^{AB}\right\}  $ are $d\times d$ mutually orthogonal maximally entangled pure
states, the upper bounds (\ref{up}) and (\ref{up2}) and the lower bound
(\ref{low}) of $N\left(  \varepsilon\right)  $ are the same and given by an
analytical expression%
\begin{equation}
N\left(  \varepsilon\right)  =S\left(  \rho^{AB}\right)  -S\left(  \rho
^{B}\right)  =S\left(  \rho^{AB}\right)  -\log d. \label{vv}%
\end{equation}
The expression (\ref{vv}) will be used when we address the entanglement charge
of some thermal states.

\textit{Entanglement charge of bipartite states.}--Consider the bipartite
quantum state $\rho^{AB}$. If it is a mixed state, it has many ensemble
decompositions \cite{book}. For example, the two-qubit state $\rho^{AB}%
=\frac{I}{2}\otimes\frac{I}{2}$ can be decomposed as an ensemble consisting of
the four computation-basis states with equal probabilities or an ensemble
consisting of the four Bell states with equal probabilities. Among all the
ensemble decompositions of $\rho^{AB}$, we can select a specific one and
define the entanglement charge of $\rho^{AB}$ as the entanglement charge of
this selected ensemble. The question is which ensemble should be selected. To
define the entanglement charge of $\rho^{AB}$, we select the ensemble
consisting of the eigenstates of $\rho^{AB}$, with the probabilities being the
corresponding eigenvalues. When the eigenvalues of $\rho^{AB}$ have no
degenerate levels, this ensemble decomposition is unique. If it is a
degenerate case, we could further require that the selected ensemble has the
maximal entanglement charge while still keep the eigenstates of $\rho^{AB}$
being mutually orthogonal.

The above defined entanglement charge $N\left(  \rho^{AB}\right)  $ of the
state $\rho^{AB}$ can be expressed as%
\begin{equation}
N\left(  \rho^{AB}\right)  =\max_{\substack{\rho^{AB}=\sum_{i}p_{i}\left\vert
\Psi_{i}\right\rangle \left\langle \Psi_{i}\right\vert ,\\\left\langle
\Psi_{i}\right\vert \left.  \Psi_{j}\right\rangle =\delta_{ij}}}\left[
N\left(  \left\{  p_{i},\left\vert \Psi_{i}\right\rangle \left\langle \Psi
_{i}\right\vert \right\}  \right)  \right]  ,
\end{equation}
where $N\left(  \left\{  p_{i},\left\vert \Psi_{i}\right\rangle \left\langle
\Psi_{i}\right\vert \right\}  \right)  $ denotes the entanglement charge of
the ensemble $\left\{  p_{i},\left\vert \Psi_{i}\right\rangle \left\langle
\Psi_{i}\right\vert \right\}  $. The reasons
are: (1) the states of the selected ensemble are mutually orthogonal,
simplifying the evaluation of the entanglement charge; (2) for thermal states,
the selected ensemble decomposition reflects the opinion that the thermal
system may be in some unknown eigenstates of the Hamiltonian; (3) for a
degenerate case, the additional requirement that the selected ensemble has the
maximal entanglement charge in all orthogonal eigenstate decompositions
represents an extreme case.

It is mentioned in the previous section that the value of the entanglement
charge of ensembles may be positive, zero or negative, so is the entanglement
charge of a state. Corresponding to the cases of ensembles, states with
positive entanglement charge are defined to have information nonlocality and
states with negative entanglement charge are said to have entanglement
nonlocality. So bipartite quantum states can be divided into three categories:
One has the information nonlocality, while the other has the entanglement
nonlocality; otherwise has neither. This appears to be a new view-angle to
understand quantum states and their nonlocalities.

The concept of information nonlocality and entanglement nonlocality introduced
here for bipartite states has a non-trivial relationship to the usual
separation and entanglement for bipartite states. On one hand, entangled
states may have information nonlocality, while unentangled states may also
have information nonlocality. For example, the two-qubit state $\rho
^{AB}=\frac{I}{2}\otimes\frac{I}{2}$ is unentangled, but its entanglement
charge $N\left(  \rho^{AB}\right)  =1$, having the information nonlocality. On
the other hand, a state having entanglement nonlocality must be entangled;
when $\rho^{AB}$ is a pure entangled state, its entanglement charge $N\left(
\rho^{AB}\right)  $ is negative, and the absolute value $\left\vert N\left(
\rho^{AB}\right)  \right\vert $ will be the quantum entropy of $\rho
^{A}=Tr_{B}\rho^{AB}$, which is the distillable entanglement of $\rho^{AB}$
\cite{book}.

\textit{Entanglement charge of thermal states.}--We have generalized the
concept of entanglement charge from ensembles to bipartite states. As an
example, we here address the entanglement charges of some thermal states. We
first consider a two-qubit system with a general XYZ interaction. The thermal
state of the system is specified by the system Hamiltonian and temperature
$T$. The system Hamiltonian reads%
\begin{equation}
H=J_{1}\sigma_{x}\otimes\sigma_{x}+J_{2}\sigma_{y}\otimes\sigma_{y}%
+J_{3}\sigma_{z}\otimes\sigma_{z},
\end{equation}
where $J_{1}$, $J_{2}$, and $J_{3}$ are real coupling parameters, and
$\sigma_{x}$, $\sigma_{y}$, and $\sigma_{z}$ are the Pauli operators. It can
be checked that the four Bell states are its eigenstates, \textit{i.e.},%
\begin{equation}
H\left\vert \Phi_{j}\right\rangle =E_{j}\left\vert \Phi_{j}\right\rangle
,\text{ }j=1,2,3,4,
\end{equation}
with%
\begin{align}
\left\vert \Phi_{1}\right\rangle  &  =\frac{1}{\sqrt{2}}\left(  \left\vert
00\right\rangle -\left\vert 11\right\rangle \right)  ,\left\vert \Phi
_{2}\right\rangle =\frac{1}{\sqrt{2}}\left(  \left\vert 00\right\rangle
+\left\vert 11\right\rangle \right)  ,\\
\left\vert \Phi_{3}\right\rangle  &  =\frac{1}{\sqrt{2}}\left(  \left\vert
01\right\rangle +\left\vert 10\right\rangle \right)  ,\left\vert \Phi
_{4}\right\rangle =\frac{1}{\sqrt{2}}\left(  \left\vert 01\right\rangle
-\left\vert 10\right\rangle \right)  ,
\end{align}
and%
\begin{align}
E_{1}  &  =-J_{1}+J_{2}+J_{3},E_{2}=+J_{1}-J_{2}+J_{3},\\
E_{3}  &  =+J_{1}+J_{2}-J_{3},E_{4}=-J_{1}-J_{2}-J_{3},
\end{align}
where $\sigma_{z}\left\vert 0\right\rangle =\left\vert 0\right\rangle $ and
$\sigma_{z}\left\vert 1\right\rangle =-\left\vert 1\right\rangle $. When the
system is in thermal equilibrium, it can be described by the Bell-diagonal
state $\rho=\sum_{j=1}^{4}p_{j}\left\vert \Phi_{j}\right\rangle \left\langle
\Phi_{j}\right\vert $, where%
\begin{equation}
p_{j}=e^{-\frac{E_{j}}{kT}}\left/  \sum_{k=1}^{4}e^{-\frac{E_{k}}{kT}}\right.
,\text{ }j=1,2,3,4.
\end{equation}
Here $k$ is the Boltzmann constant.

The entanglement charge $N\left(  \rho\right)  $ of the above thermal state
$\rho$
can be calculated as follows. When the eigenvalues of $\rho$ are not
degenerate, $\rho$ has the unique eigenstates decomposition $\left\{
p_{j},\left\vert \Phi_{j}\right\rangle \left\langle \Phi_{j}\right\vert
\right\}  _{j=1}^{4}$. Since the eigenstates of $\rho$ are mutually orthogonal
maximally entangled states, an analytical expression for entanglement charge
$N\left(  \rho\right)  $ is obtained from Eq.(\ref{vv}),%
\begin{equation}
N\left(  \rho\right)  =S\left(  \rho\right)  -1, \label{aaa}%
\end{equation}
where $S\left(  \rho\right)  =-\sum_{j=1}^{4}p_{j}\log_{2}p_{j}$ is the
entropy of the system. We note that even when the eigenvalues of $\rho$ are
degenerate, the expression (\ref{aaa}) is still valid for $N\left(
\rho\right)  $ because it reaches the maximal entanglement charge of all
ensemble decompositions of $\rho$, which can be seen for Eq. (\ref{up}). From
Eq. (\ref{aaa}), we find that the entanglement charge is just a shifted
entropy in this case, however the meanings of the entanglement charge
$N\left(  \rho\right)  $ and the entropy $S\left(  \rho\right)  $ are entirely different.

It is not hard to find that the entanglement charge $N\left(  \rho\right)  $
of this thermal state ranges between $-1$ and $1$. The exact value of
$N\left(  \rho\right)  $ depends on the temperature and the system coupling
parameters. Clearly, the states with positive $N\left(  \rho\right)  $ are
different from those with negative $N\left(  \rho\right)  $ in the sense that
they have different kinds of nonlocalities, so we can investigate the change
of the nonlocality properties due to the change of the temperature and the
system coupling parameters. In the following we will consider three kinds of
models to explore this nonlocality property change. As a comparison, the
entanglement of the states will also be given, with the concurrence being
chosen as the entanglement measure. The considered thermal state $\rho
=\sum_{j=1}^{4}p_{j}\left\vert \Phi_{j}\right\rangle \left\langle \Phi
_{j}\right\vert $ is a Bell-diagonal state, whose concurrence is given by
\cite{con}%
\begin{equation}
C\left(  \rho\right)  =\max\left\{  1,2p_{1},2p_{2},2p_{3},2p_{4}\right\}  -1.
\end{equation}
\begin{figure}
[ptb]
\begin{center}
\includegraphics[
height=2.0358in,
width=3.039in
]%
{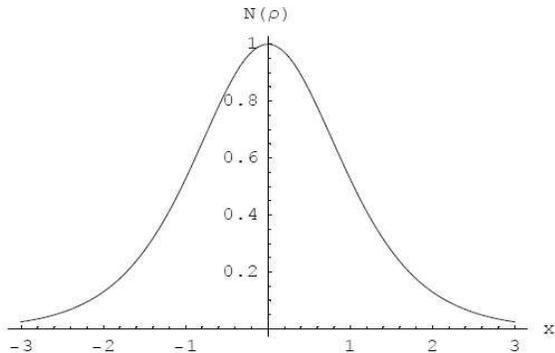}%
\caption{The entanglement charge $N\left(  \rho\right)  $ of the thermal state
of $\rho$ where the system has an Ising Hamiltonian $H=J_{1}\sigma_{x}%
\otimes\sigma_{x}$. The entanglement charge $N\left(  \rho\right)  $ is
determined by one parameter $x=J_{1}/kT$. }%
\label{figfig1}%
\end{center}
\end{figure}

First, we consider the Ising model. In this case $J_{2}=J_{3}=0$, so the
eigenvalues of the Hamiltonian are $E_{1}=-J_{1}$, $E_{2}=J_{1}$, $E_{3}%
=J_{1}$, and $E_{4}=-J_{1}$. The entanglement charge $N\left(  \rho\right)  $
of the thermal state depends only on the parameter $x=J_{1}/kT$. In FIG. 1 we
plot $N\left(  \rho\right)  $ as a function of $x$. It can be seen that the
entanglement charge is always positive. This can be understood from the fact
that $p_{1}=p_{4}$, $p_{2}=p_{3}$ and $p_{1}+p_{2}+p_{3}+p_{4}=1$, which leads
to $p_{1}+p_{2}=p_{3}+p_{4}=1/2$ and a positive
\begin{equation}
N\left(  \rho\right)  =-2p_{1}\log_{2}\left(  2p_{1}\right)  -2p_{2}\log
_{2}\left(  2p_{2}\right)  . \label{xx}%
\end{equation}
The probabilities $p_{1}$ and $p_{2}$ will be exchanged when we change $x$ to
$-x$, so Eq. (\ref{xx}) indicates $N\left(  \rho\right)  $ is an even function
of $x$. The concurrence of the thermal state $\rho$ in this case is always
zero since no $p_{i}$ is bigger than $1/2$.%

\begin{figure}
[ptb]
\begin{center}
\includegraphics[
height=1.9597in,
width=3.039in
]%
{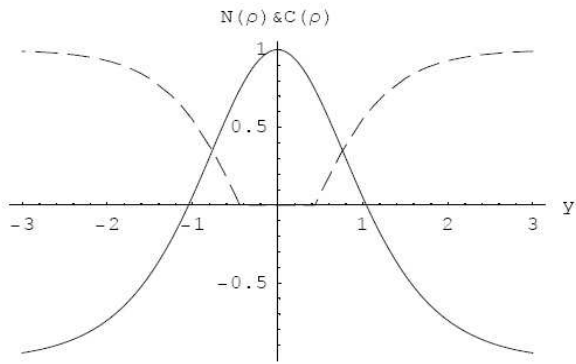}%
\caption{The entanglement charge $N\left(  \rho\right)  $ (the solid line) and
the concurrence $C\left(  \rho\right)  $ (the dashed line) of the thermal
state of $\rho$ where the system has an XX Hamiltonian $H=J_{1}\sigma
_{x}\otimes\sigma_{x}+J_{1}\sigma_{y}\otimes\sigma_{y}$. The entanglement
charge $N\left(  \rho\right)  $ and the concurrence $C\left(  \rho\right)  $
are determined by one parameter $y=J_{1}/kT$. }%
\label{figfig2}%
\end{center}
\end{figure}

Secondly, we consider the XX model. In this case $J_{2}=J_{1}$, $J_{3}=0$, so
the eigenvalues of the Hamiltonian are $E_{1}=0$, $E_{2}=0$, $E_{3}=2J_{1}$,
and $E_{4}=-2J_{1}$. The entanglement charge $N\left(  \rho\right)  $ and
concurrence $C\left(  \rho\right)  $ of the thermal state depend only on
$y=J_{1}/kT$. In FIG. 2 we plot $N\left(  \rho\right)  $ (the solid line) and
the concurrence $C\left(  \rho\right)  $ (the dashed line) as a function of
$y$. The entanglement charge $N\left(  \rho\right)  $ is an even function of
$y$ can be understood from the fact that $p_{3}$ and $p_{4}$ will be exchanged
when we change $y$ to $-y$. It can be seen that when $\left\vert y\right\vert
$ is small the thermal state has information nonlocality (positive
entanglement charge) and when $\left\vert y\right\vert $ is large the thermal
state has entanglement nonlocality (negative entanglement charge). This can be
understood as follows. When $\left\vert y\right\vert \rightarrow0$, it can be
regarded as the temperature $T\rightarrow\infty$, which leads the system to be
in one of the four Bell states with an equal probability $1/4$ and $N\left(
\rho\right)  =1$. When $\left\vert y\right\vert \rightarrow\infty$, it can be
regarded as the temperature $T\rightarrow0$, which leads the system to be in
the ground state $\left\vert \Phi_{4}\right\rangle $ (when $J_{1}>0$) or
$\left\vert \Phi_{3}\right\rangle $ (when $J_{1}<0$), and $N\left(
\rho\right)  =-1$. So the thermal state with information nonlocality will be
changed to the state with entanglement nonlocality when the temperature is
decreased (i.e., $\left\vert y\right\vert $ is increased). It can be seen that
when the state $\rho$ has information nonlocality (positive $N\left(
\rho\right)  $), there are regions of y where the state $\rho$ is entangled
(positive $C\left(  \rho\right)  $) and there are also regions of y where the
state $\rho$ is not entangled, which indicates the information nonlocality has
no direct relation to quantum entanglement.%

\begin{figure}
[ptb]
\begin{center}
\includegraphics[
height=1.9778in,
width=3.039in
]%
{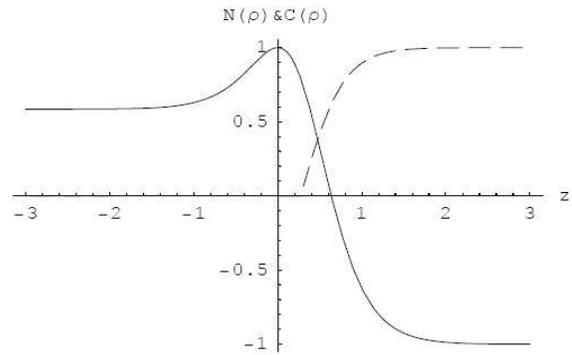}%
\caption{The entanglement charge $N\left(  \rho\right)  $ (the solid line) and
the concurrence $C\left(  \rho\right)  $ (the dashed line) of the thermal
state of $\rho$ where the system has an Heisenberg exchange Hamiltonian
$H=J_{1}\sigma_{x}\otimes\sigma_{x}+J_{1}\sigma_{y}\otimes\sigma_{y}%
+J_{1}\sigma_{z}\otimes\sigma_{z}$. The entanglement charge $N\left(
\rho\right)  $ and the concurrence $C\left(  \rho\right)  $ are determined by
one parameter $z=J_{1}/kT$. }%
\label{figfig3}%
\end{center}
\end{figure}

Finally, we discuss the Heisenberg model. In this case $J_{2}=J_{3}=J_{1}$, so
the eigenvalues of the Hamiltonian are $E_{1}=J_{1}$, $E_{2}=J_{1}$,
$E_{3}=J_{1}$, and $E_{4}=-3J_{1}$. The entanglement charge $N\left(
\rho\right)  $ and the concurrence $C\left(  \rho\right)  $ of the thermal
state only depend on $z=J_{1}/kT$. In FIG. 3 we plot $N\left(  \rho\right)  $
(the solid line) and the concurrence $C\left(  \rho\right)  $ (the dashed
line) as a function of $z$. It can be seen that $N\left(  \rho\right)  $ is an
asymmetrical function, which is different from the above Ising and XX models.
When $z\rightarrow\infty$, it can be regarded as the temperature
$T\rightarrow0$ in the antiferromagnetic case(i.e., $J_{1}>0$), which leads
the system to be in the ground state $\left\vert \Phi_{4}\right\rangle $ and
$N\left(  \rho\right)  =-1$. When $z\rightarrow-\infty$, it can be regarded as
the temperature $T\rightarrow0$ in the ferromagnetic case (i.e., $J_{1}<0$),
which leads the system to be in the states $\left\vert \Phi_{1}\right\rangle
$, $\left\vert \Phi_{2}\right\rangle $, and $\left\vert \Phi_{3}\right\rangle
$ with an equal probability $1/3$ and $N\left(  \rho\right)  =\log
_{2}3-1\thickapprox0.58$. It can be seen that in the ferromagnetic case
($z<0$) the thermal state has no entanglement while it has the information
nonlocality. And in the antiferromagnetic case ($z>0$), the region of z where
the state has entanglement nonlocality (negative $N\left(  \rho\right)  $) is
contained in the region of z where the state is entangled (positive $C\left(
\rho\right)  $), which just manifests the fact that only entangled states may
have entanglement nonlocality.

In the above two-qubit Heisenberg model, the two-qubit thermal state is in the
form
\begin{equation}
\rho=p_{1}\left(  \left\vert \Phi_{1}\right\rangle \left\langle \Phi
_{1}\right\vert +\left\vert \Phi_{2}\right\rangle \left\langle \Phi
_{2}\right\vert +\left\vert \Phi_{3}\right\rangle \left\langle \Phi
_{3}\right\vert \right)  +p_{4}\left\vert \Phi_{4}\right\rangle \left\langle
\Phi_{4}\right\vert , \label{f}%
\end{equation}
which is a linear combination of the projector to the triplet space and the
projector to the singlet space. This is a common feature of the two-qubit
states that have the $SU\left(  2\right)  $ symmetry \cite{su2,ww}. Now we
consider a qubit ring consisting of $M$ qubits with the Hamiltonian%
\begin{equation}
H=J\sum_{i=1}^{M-1}\vec{\sigma}_{i}\cdot\vec{\sigma}_{i+1}+J\vec{\sigma}%
_{M}\cdot\vec{\sigma}_{1},
\end{equation}
where $\vec{\sigma}_{i}=\left(  \sigma_{ix},\sigma_{iy},\sigma_{iz}\right)  $
is the vector of Pauli operators. The ring has the translation symmetry, so
any two adjacent qubits of the ring will have the same thermal state described
by its reduced density matrix. This thermal state of two adjacent qubits of
the ring will also have the $SU\left(  2\right)  $ symmetry,
and thus it can be written in the form given by Eq. (\ref{f}). When we
consider the entanglement charge of the thermal state of the two adjacent
qubits of the ring, some features of the above two-qubit Heisenberg model can
be obtained since the states in these two cases have the same form as Eq.
(\ref{f}). For a ferromagnetic ring (i.e., the effective coupling $J<0$), the
thermal state of the two adjacent qubits cannot have the entanglement
nonlocality since it has no entanglement at any temperature \cite{wz}, which
is the same as that for the above two-qubit Heisenberg model. For an
antiferromagnetic ring (i.e., the effective coupling $J>0$), the entanglement
charge of the thermal state of the two adjacent qubits cannot reach $-1$ even
when the temperature $T$ goes to zero, which is different from that for the
above two-qubit Heisenberg model. Though the thermal state of two adjacent
qubits of the ring has the form in Eq. (\ref{f}), an explicit expression of
the state parameter $p_{1}$ for a general $M$ (the number of the qubits in the
ring) is to be derived, which seems quite hard and may be handled in future.

\textit{Summary.}--We have generalized the concept of entanglement charge of
ensembles to single bipartite quantum states. According to their entanglement
charges, bipartite quantum states can be divided into three categories that
have the information nonlocality, the entanglement nonlocality, and neither.
This is a new view-angle to understand quantum states and their nonlocalities.
As an example we have addressed entanglement charges of some thermal states of
two-qubit systems. We have found that for some simple models, the thermal
states with information nonlocality can be changed to states with entanglement
nonlocality by decreasing the temperature. The present work is expected to
evoke more profound understandings of nonlocalities in quantum states.

\textit{Acknowledgments.}--We thank helpful discussions with Q.-H. Wang. This
work was supported by NSF-China (Nos. 60878059, 10947147, 10905016),
NSF-Fujian (No2010J01002), the fund of Hebei Normal University, the RGC grant
of Hong Kong under No. HKU7044/08P, and the State Key Program for Basic
Research of China (No.~2006CB921800).

\end{document}